\documentclass[final,5p,times,twocolumn]{elsarticle}
\journal{Nuclear Instruments and Methods in Physics Research A}
\usepackage{amsmath}

\usepackage[english]{babel}
\usepackage[T1]{fontenc}
\usepackage[utf8]{inputenc}
\usepackage{color}
\usepackage{graphicx} 
\usepackage{pgfplots}
\usepackage{tikz}
\usetikzlibrary{3d}

\def\Emax{E_{\mathrm{max}}}
\def\parton#1{\left({{#1}}\right)}
\def\parqua#1{\left[{{#1}}\right]}
\def\Ecal{{\cal E}}
\def\derp#1#2{{\partial{#1} \over \partial{#2}}}
\def\aladyn{\texttt{ALaDyn} }
\begin{document}

\title{Rise time of proton cut-off energy in 2D and 3D PIC simulations}

\author[iran]{Javad~Babaei}
\author[cnr]{Leonida~Antonio~Gizzi}
\author[unibo]{Pasquale~Londrillo}
\author[iran]{Saeed~Mirzanejad}
\author[unibo]{Tiziano~Rovelli}
\author[unibo]{Stefano~Sinigardi\corref{cor1}}
\ead{Stefano.Sinigardi@bo.infn.it}
\cortext[cor1]{Corresponding author}
\author[unibo]{Giorgio~Turchetti}

\address[iran]{Department of Physics, Faculty of Basic Sciences, University of Mazandaran, P. O. Box 47415-416, Babolsar, Iran}
\address[cnr]{ILIL, Istituto Nazionale di Ottica, CNR Pisa \& INFN Sezione di Pisa, Italy}
\address[unibo]{Dipartimento di Fisica e Astronomia, Universit\`a di Bologna and INFN Sezione di Bologna, Via Irnerio 46, I-40126 Bologna (BO), Italy}

\date{\today}
\begin{keyword}
laser driven ion acceleration \sep particle-in-cell simulations
\end{keyword}

\begin{abstract}

The Target Normal Sheath Acceleration (TNSA) regime for proton acceleration by laser pulses is
experimentally consolidated and fairly well understood.
However, uncertainties remain in the analysis of particle-in-cell (PIC) simulation results.

The energy spectrum is exponential with a cut-off, but
the maximum energy depends on the simulation time, following different
laws in two and three dimensional (2D, 3D) PIC simulations, so that the determination
of an asymptotic value has some arbitrariness. 

We propose two empirical laws for rise time
of the cut-off energy in 2D and 3D PIC simulations, suggested by a model in
which the proton acceleration is due to a surface charge
distribution on the target rear side. 
The kinetic energy of the protons that we obtain follows two distinct laws, which
appear to be nicely satisfied by PIC simulations.
The laws depend on two parameters: the scaling time,
at which the energy starts to rise, and the asymptotic
cut-off energy. 

The values of the cut-off energy, obtained by fitting the
2D and 3D simulations for the same target and laser pulse,
are comparable. This suggests that parametric scans can be performed with 2D simulations,
since 3D ones are computationally very expensive.
In this paper, the simulations are carried out for $a_0=3$ with the
PIC code \aladyn by changing the target thickness $L$
and the incidence angle $\alpha$. A monotonic dependence, on $L$
for normal incidence and on $\alpha$ for fixed $L$,
is found, as in the experimental results for high temporal contrast pulses.
\end{abstract}


\maketitle
 
\section{Introduction}
\label{intro}
The acceleration of protons by intense laser pulses is still the subject of 
active experimental investigation. The most consolidated regime is the TNSA,
where the electrons, heated by laser, diffuse and leave the target creating
an electric field which accelerates the surface protons present in the contaminants.
The comparison with current PIC simulations is still affected by uncertainties.
Indeed the energy spectra are found to be exponential with a cut-off 
\[
  \begin{cases}
    \displaystyle dN/dE = (\Emax/T)\,\, e^{-E/T} &\mathrm{for}\; E<\Emax \\
    \displaystyle dN/dE = 0                      &\mathrm{for}\; E>\Emax
  \end{cases}
\]
but the cut-off energy $\Emax$ and the average energy value $T$ (proton temperature)
depend on time. In 2D PIC TNSA simulations, a monotonic rise of
$\Emax$ with time is observed whereas in 3D
a slow trend towards a possible saturation to an asymptotic value is usually observed.
As a consequence, a comparison of 2D and 3D simulations is difficult,
since the laws of the cut-off energy rise with time $\Emax(t)$ appear
to be different. 

In this paper we try to give a phenomenological answer to this
question, by proposing two empirical laws for $\Emax(t)$, 
suggested by a model firstly proposed by Schreiber et al. \cite{Schreiber}, to describe the
dependence of the cut-off energy from the laser pulse duration.
This model assumes that the hot electron cloud leaves the rear side of the
target, creating a surface density of positive charge, whose electric field
accelerates the protons belonging to the contaminants.
We have considered a 2D model in which the surface charge is on a strip with infinite
length and height $2R$, with $R$ corresponding to the laser waist, and a 3D model in which the surface charge
is located on a disc of radius $R$.
In our model, the laser is assumed to have normal incidence on the target and 
in figure 1 we sketch the geometric configurations.

%
%
%

The numerical analysis presented here refers to a laser pulse with
$\tau=40$ fs
and $a_0=3$. This choice was made because, recently, systematic experiments
with such a laser pulse were carried out at ILIL in Pisa \cite{Gizzi}.
Besides, several experiments with similar parameters,
which ensure the acceleration regime is TNSA, are present in the literature.
For an overview on the physics of the proton acceleration by high intensity
lasers and related experiments, we refer to recent reviews \cite{Borghesi,Macchi_2_RMP,Daido}.
In the considered intensity range, experimental results concerning the dependence on the target thickness, 
the incidence angle and the temporal contrast are described in many papers \cite{Fritzler,Ceccotti,Zeil,Spencer,Neely,Yogo,Flacco}.
When the contrast is very high, the cut-off energy varies monotonically
with target thickness and if the contrast were infinite this behaviour
would be observed, until the radiation pressure becomes dominant by approaching 
the relativistic transparency limit.
When the contrast is finite, as in experiments, a maximum in the cut-off
energy $\Emax$ is reached at a certain minimum thickness. 
By further reducing thickness, a rapid decrease to zero of $\Emax$
is observed, due to the increasing damage on the foil induced
by the prepulse.
%
%
A significant dependence on the incidence angle
is also observed and typically the proton cut-off energy
increases with the angle up to a maximum value, because the electrons are heated more efficiently \cite{Macchi_2_RMP}.
%
%
%

In our model, the preplasma is neglected (the temporal contrast is assumed as infinite).
Because of this choice, Amplified Spontaneous Emission (ASE) prepulse is not permitted. On the other hand, prepulse coming 
from compression artefacts (ps time scale) can be tolerated when comparing our simulation results with experiments,
as long as the plasma preformed on the illuminated side of the target has a scale length much shorter than the laser wavelength.

The 2D and 3D simulations were carried out with the \aladyn code
\cite{ALaDyn_GPL_20160412} and the asymptotic cut-off energy $E_\infty$ was determined by a best-fit procedure
on its time dependence, following the laws obtained from the
electrostatic model, which just depend on two parameters: the
asymptotic cut-off energy $E_\infty$ and the rise time $t^*$, which is the time when the energy starts to rise. 


Beyond the good agreement of the asymptotic cut-off
energies obtained from 2D and 3D 
simulations, the monotonic dependence on the incidence angle and the target
thickness was found in qualitative agreement with the experimental
results for high contrast pulses. 

In our 3D simulations, the transverse section of the computational box is the same
as the target, whose extension is comparable with the focal spot (four times bigger) measured by
the waist. As a consequence, a leakage of electrons from the
computational box occurs and when the fraction of lost electrons becomes
appreciable, typically for $ct$ significantly above $100 \mu$m, the simulation looses reliability.
That is why we stop our analysis at this time. Increasing the box size would enable us to go further,
but without adding any insightful detail.

Our method allows us to limit the simulation even to  
$ct= 60 \sim 80\, \mu$m using small boxes, since the results are already stable 
and comparable with 3D results. 
Here we present the results for a single laser pulse and various
target thicknesses, to assess the validity of our model, even though 
we have started a more extensive exploration by varying the laser
duration, its intensity and the metal target electron density.
A detailed analysis of the dependence of $E_\infty$
and $t^*$ on laser and target parameters will give us a better
insight, but, from the encouraging results obtained so far, we can
conclude that the simple method we propose here appears to be adequate 
to extract the asymptotic cut-off energy from PIC simulations.

\section{The 3D case}
Starting from the 3D case and considering a laser pulse which propagates
along the $z$ axis, we choose an electrostatic potential which vanishes at
$z=0$, where the surface charge (density $\sigma$) is located. 
This potential is given by
\[ V(\zeta)= 2\pi R\,\sigma\,\Bigl( \sqrt{1+\zeta^2} -\zeta -1 \Bigr ) \qquad \qquad
\zeta={z\over R} \]
Asymptotically, for $z\to \infty$, it behaves as $V = Q/z$, where $Q=\pi R^2 \sigma$
is the charge on the disc. A particle initially at rest accelerates
and the law of motion is obtained from energy conservation. Since $V(0)=0$,
we have 
\[ m{v^2\over 2} +eV(z)=0 \qquad \quad v=\dot z \]
Letting $v_\infty=\dot z(\infty)$, the kinetic energy
of the particle, after integrating the equation of motion, is
\[ E(t) \simeq E_\infty\parton{ 1-{t^*\over t}}^2 \qquad \quad t>t^*
= {R\over 4v_\infty}\]
where
\[ E_\infty= m{v_\infty^2\over 2}=2\pi eR\sigma \]
Since this is an asymptotic law, we may assume that $E(t)=0$ for $t<t^*$.
Notice that $E$ is the highest energy reached at time $t$, namely $E=\Emax$. 
%
\begin{figure}
\centering
\includegraphics[width=0.40 \textwidth]{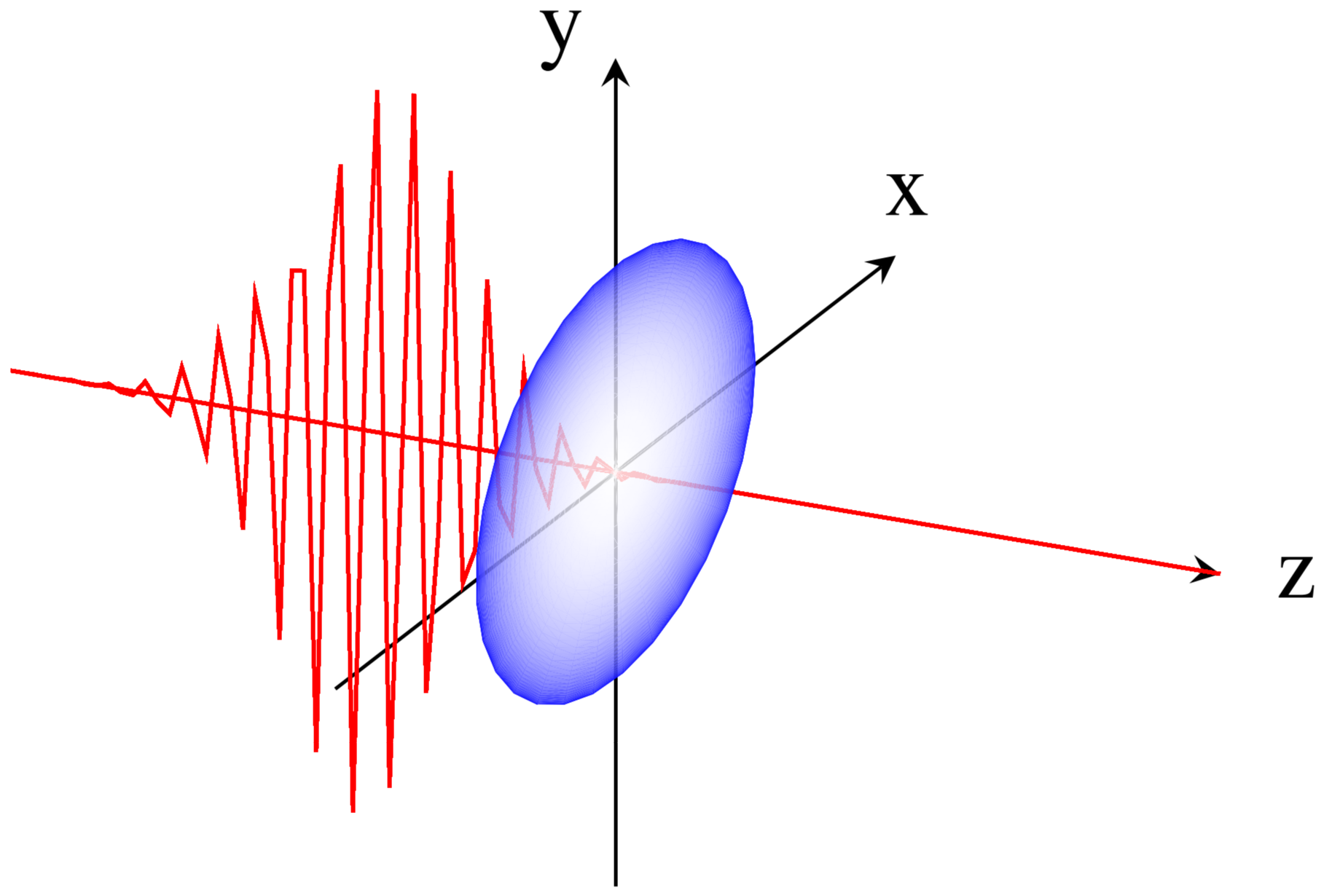} \\
\vspace{3mm}
\includegraphics[width=0.40 \textwidth]{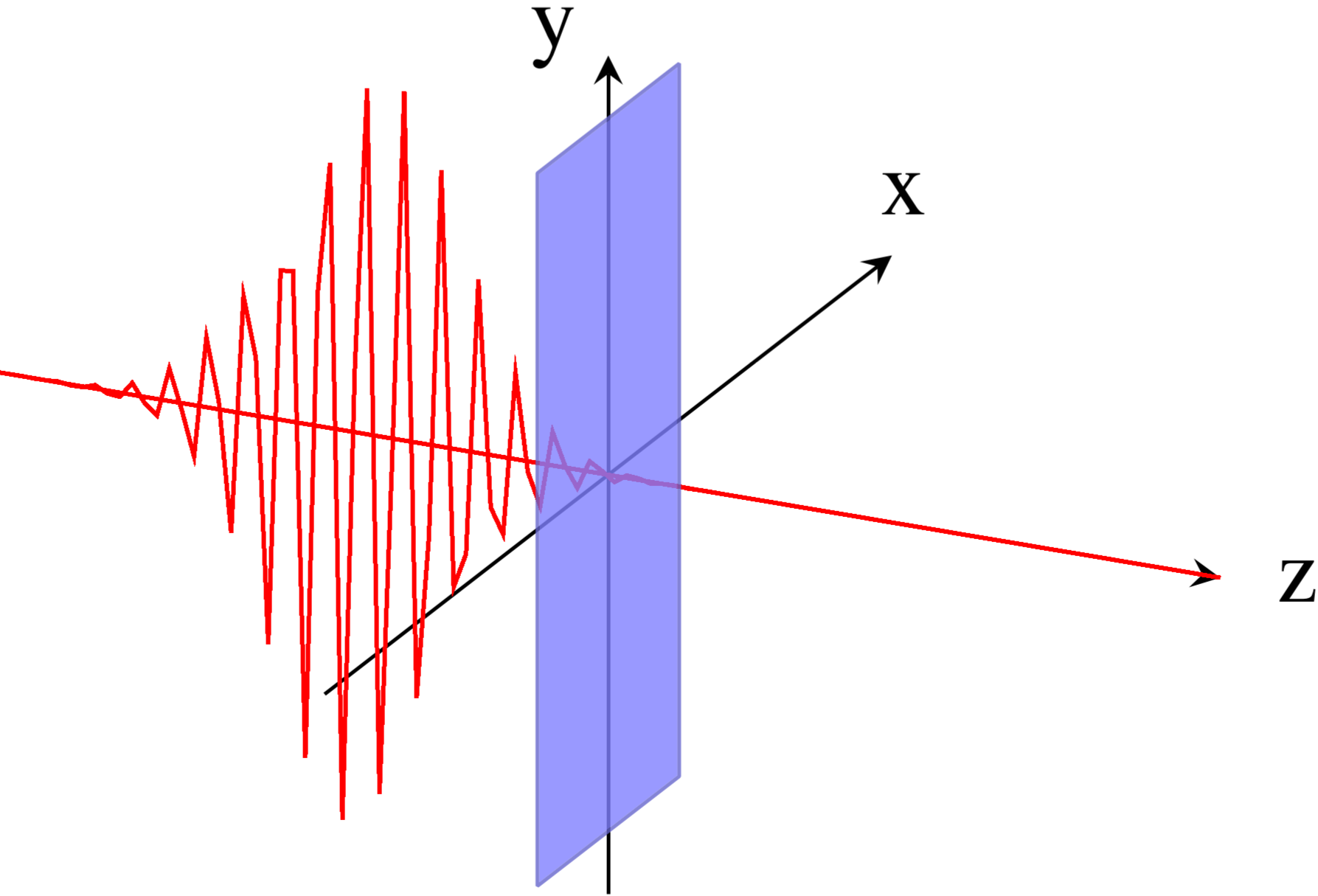} 
\caption{Schematic representation of the configurations used to compute the accelerating field: 3D (above) and 2D (below).}
\label{fig:1}
\end{figure}

\section{The 2D case}
In this case we have a infinite strip along the $y$ axis with
uniform charge density on $-R<x<R$. A potential that vanishes at $z=0$
is given by 
\[
  \begin{split}
    \displaystyle V(z) &= 4R\sigma \parton{-\zeta \arctan {1\over \zeta} +\log{1\over\sqrt{1+\zeta^2}} }\\
    \displaystyle &\simeq -4R\sigma\, \log(1+\zeta)
  \end{split}
\]
%
where we defined \(\zeta=z/R\).
To obtain this result, it is simpler to compute first the electric field
$\Ecal_x= 4\sigma\,\, \arctan(1/\zeta)$, whose asymptotic behaviour
is $4\sigma/\zeta$. As a consequence, a potential having this asymptotic 
behaviour and which vanishes at the origin is 
$\hat{V}\simeq -4R\sigma\, \log(1+\zeta)$. The potential in this case diverges
logarithmically and consequently the particle accelerates indefinitely.
We approximate the potential energy with
\[ e\hat{V}(z)= - E_\infty\, \log(1+\zeta) \qquad \qquad
E_\infty\equiv m{v_\infty^2\over 2}= 4eR\sigma \]
We may then easily solve the equations of motion from energy conservation,
assuming the proton initially at rest in the origin as for the 3D case.
The result is (see Appendix for more details)
\[ E(t)= E_\infty \log\parton{t\over t^*} \qquad \qquad
t\geq t^*= {R\over v_\infty} \]
Again, since this is an asymptotic law, we may assume that
$E(t)=0$ for $t<t^*$. 
\section{Comparison with PIC simulations }
Even though the models we propose are very simple, we tried to see whether the
predicted asymptotic laws for $E(t)$ hold for PIC simulations.
The answer is positive, at least for targets consisting of a uniform
foil whose thickness is in the micrometer range, covered by a thin layer of contaminants.
For this type of targets, the fits, both for 2D and 3D PIC simulations, 
are surprisingly accurate. However, the asymptotic
energy $E_\infty$ and the time scale $t^*$ in 2D and 3D must be considered fitting
parameters, even though the results we obtain have the correct order of magnitude
with respect to the theoretical results.

The law to be fitted for 2D simulations is
\[
\left\{
  \begin{array}{ll}
    \displaystyle \Emax^{(2D)}(ct) = 0                                   &\mathrm{for}\; t<t^{*(2D)} \\
    \displaystyle \Emax^{(2D)}(ct) = E^{(2D)}_\infty\,\log{ct\over ct^*} &\mathrm{for}\; t>t^{*(2D)}
  \end{array}
\right.
\]
We perform a linear fit by defining $y=E$ and $x=\log ct$, so that the previous law
becomes
\[ y=a+bx \qquad \qquad \qquad E^{(2D)}_\infty=b \qquad ct^{*(2D)}=e^{-a/b} \]

The law to be fitted for 3D simulations is 
\[
\left\{
  \begin{array}{ll}
    \displaystyle \Emax^{(3D)}(ct) = 0                                                   &\mathrm{for}\; t<t^{*(3D)} \\
    \displaystyle \Emax^{(3D)}(ct) = E^{(3D)}_\infty\,\parton{ 1-{ct^{*(3D)}\over ct}}^2 &\mathrm{for}\; t>t^{*(3D)}
  \end{array}
\right.
\]
We can perform a linear fit by defining $y=\sqrt{E}$ and $x=1/ct$, so that the previous law
becomes
\[ y=a+bx \qquad \qquad \qquad E^{(3D)}_\infty=a^2 \qquad ct^{*(3D)}= -{b\over a} \]

\section{Results for 2D simulations }
We have considered the following model: the laser pulse has wavelength
$\lambda=0.8 \,\mu$m, intensity $I=2\,\cdot\,10^{19} $ W/cm$^2$, waist
6.2 $\mu$m, P-polarization and its duration is 40 fs. 
The corresponding normalized vector potential is 
$a_0=3$. The target is a uniform Al foil of thickness $L$ varying between
0.5 and 8 $\mu$m, having a layer of hydrogen on the rear (non illuminated) side, 
with fixed thickness $0.08 \mu$m.

%
%
%
%
%
\begin{figure}
\centering
\includegraphics[width=0.45 \textwidth]{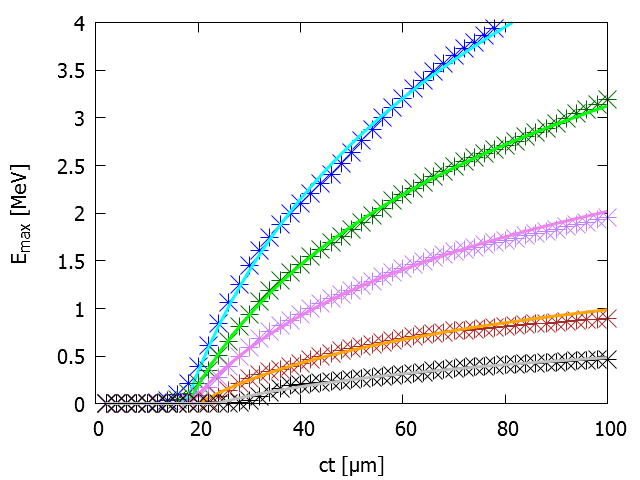}
\includegraphics[width=0.45 \textwidth]{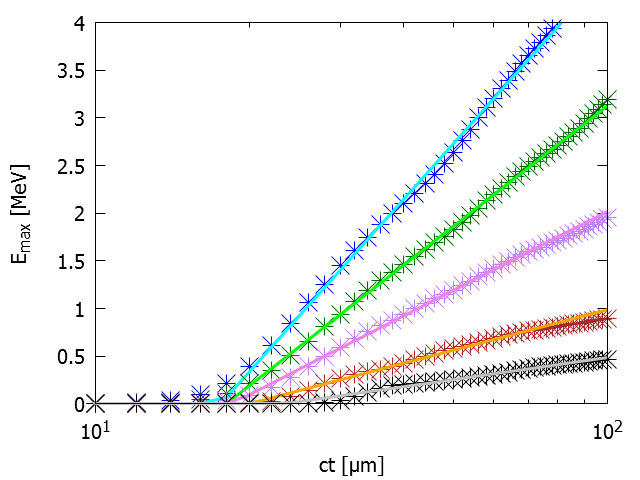} 
\caption{Above: cut-off energy $\Emax$ versus $ct$ 
in the range $10\leq ct\leq 100 \,\mu$m obtained
from a PIC simulation (stars) and comparison with the fit (continuous line)
for targets of various thicknesses $L$: 
blue (cyan) $L=0.5 \,\mu$m, dark green (green) $L=1 \,\mu$m, purple (violet) $L=2 \,\mu$m,
brown (orange) $L=4 \,\mu$m, black (grey) $L= \,8\mu$m.
Below: the same as the left panel but in a logarithmic scale for $ct$ which clearly shows the linearity and
the accuracy of the fit.}
\label{fig:2} 
\end{figure} 
\begin{figure}
\centering
\includegraphics[width=0.45 \textwidth]{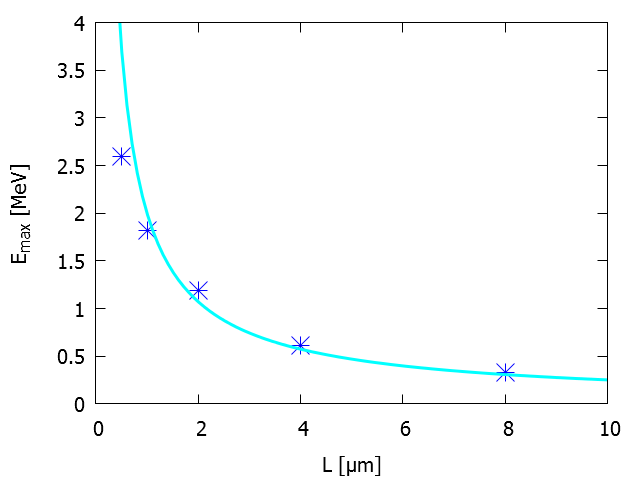}
\caption{Comparison of the extrapolated cut-off energy for 2D PIC simulations (blue stars)
for different target thicknesses $L=0.5, 1,2,4,8 \mu$m
and a fit with the curve $E_{\mathrm{max}}=1/L^{0.9}$ (cyan line).}
\label{fig:3} 
\end{figure}
%

The ionization level is Al$^{9+}$ and H$^{+}$ and it is fixed throughout the simulation. The electron densities have been chosen as
$n_e^{Al}=100 \,n_c$ and $n_e^H=10\,n_c$. 
For a Al foil, whose thickness is in the $[0.5,8]\,\mu$m range, we expect
that the process is dominated by TNSA (we are well beyond the transparency limit). 
The collisional models have been neglected in our simulations.



\begin{figure}
\centering
\includegraphics[width=0.45 \textwidth]{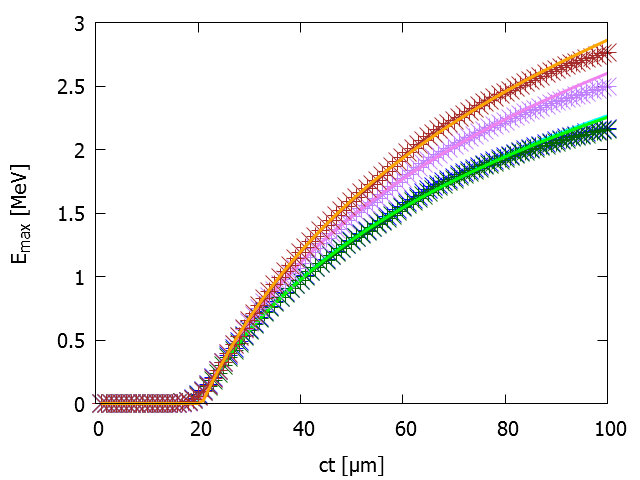}
\includegraphics[width=0.45 \textwidth]{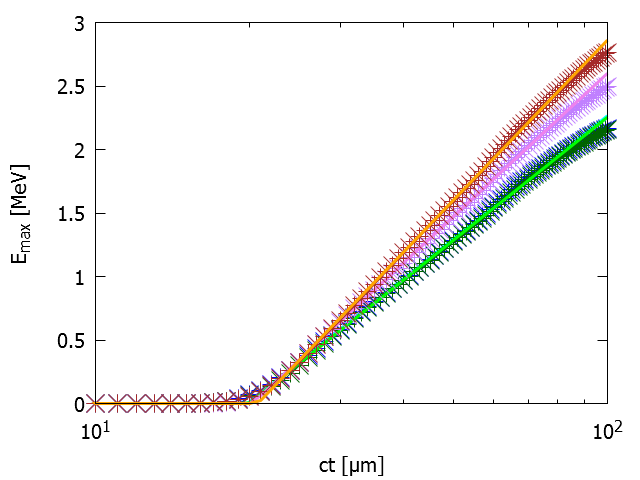}
\caption{Above: comparison of the 2D PIC solution with
a small incidence angle $\alpha$.
The figure shows $\Emax$ versus $ct$, the stars corresponding to the PIC simulation and
the curves to the fit for various angles: $\alpha=5^{\circ}$ dark green (green), $\alpha=10^{\circ}$
purple (violet) and $\alpha=15^{\circ}$ brown(orange).
Below: the same data are plotted as with a logarithmic scale for $ct$,
which shows how the data stay on a line and
the accuracy of the linear fit, see table \ref{tab:3}.}
\label{fig:4} 
\end{figure}

In figure \ref{fig:2} we show the results obtained from 2D simulations for $0.5 \mu\mathrm{m} \leq L\leq 8\mu\mathrm{m}$,
by plotting $\Emax(ct)$ in a linear and a logarithmic scale for $ct$ with the corresponding fits.
Initially, the time at which the energy starts to rise is almost independent from the thickness
$ct^*\simeq 20\, \mu$m.
In table \ref{tab:1} we quote the results of the fit: we notice that $E_\infty^{(2D)}\simeq E(ct=50)$.
In figure \ref{fig:3} we resume the dependence of the cut-off energy on the thickness.
In figure \ref{fig:4} we show the results of 2D simulations obtained when the incidence angle is small
but different from zero: the logarithmic growth in $ct$ is still present and the linear fits are quite good, 
see also table \ref{tab:2}.
%
%
%
\section{Results for 3D simulations }
\begin{figure}
\centering
\includegraphics[width=0.45 \textwidth]{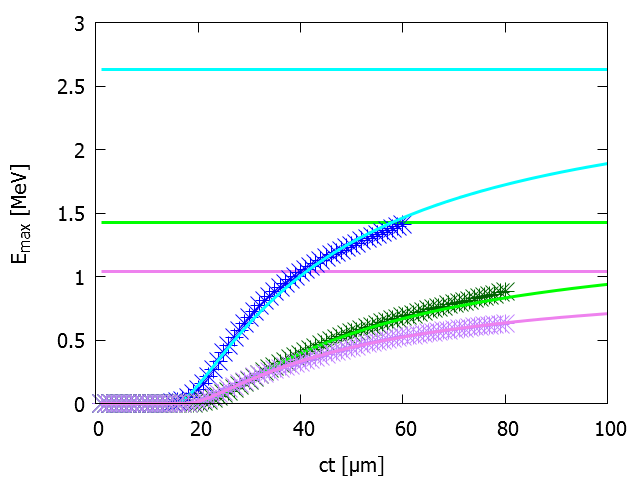}
\includegraphics[width=0.45 \textwidth]{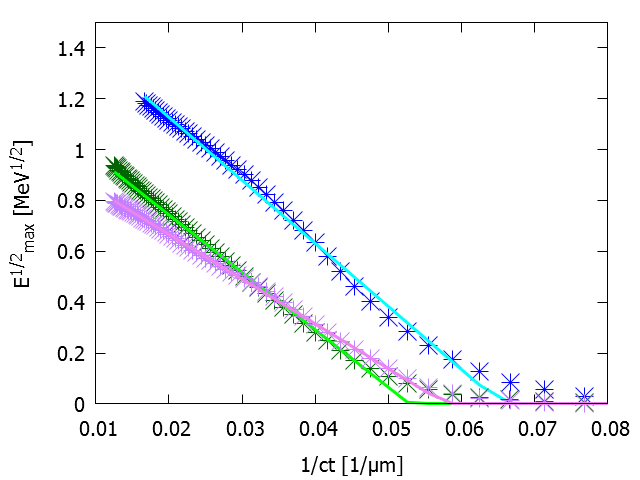}
\caption{
Above: results for a 3D PIC simulation for $\Emax$ versus $ct$ (stars)
compared with the linear fit of $\sqrt{\Emax}$ as a function of $1/ct$ (continuous
lines, the asymptotic values $E_\infty^{(3D)}$ are also shown), for different target thickness:
$L=0.5 \mu$m blue (cyan), $L=1 \mu$m dark green (green) and to $L=2 \mu$m purple (violet).
Below: Plot of $\sqrt{E_{\mathrm{max}}}$ versus $1/ct$ which
shows their linearity, with the corresponding linear fit.
}
\label{fig:5} 
\end{figure}

We present now the results for some 3D simulations, precisely with $L=0.5,1,2 \, \mu$m.
In figure \ref{fig:5} we show the curves corresponding to a linear fit to $\sqrt{E(t)}$
versus $1/ct$. The asymptotic values $E_{\infty}^{(3D)}$ and the fitting curves up to
$ct=100 \, \mu$m are shown in the left panel figure \ref{fig:5}.
%

We notice that, even though the extrapolated data from the 2D and 3D simulations are
not the same, the correspondence is quite reasonable. In table \ref{tab:1} the numerical results are quoted
and in any case the discrepancy does not exceed 20\%. 
We may observe that the energy for $ct=50 \,\mu$m in the 2D simulation is very close to the
extrapolated value, due to the logarithmic growth. In table \ref{tab:2} we report the numerical results about
the $E_\infty$ obtained for three different incidence angles $\alpha=5^{\circ},10^{\circ},15^{\circ}$
and target thickness $L=2 \mu$m. In 3D at $ct=50 \,\mu$m the energy value is
less than one half of the extrapolated value $E_\infty$ due to the slower rise, see table \ref{tab:3}. 
In this case there is an asymptotic limit, which is reached quite far, when $ct > 200 \mu$m. 
Such a large value is computationally too expensive to be attained.

\begin{figure}
\centering
\includegraphics[width=0.45 \textwidth]{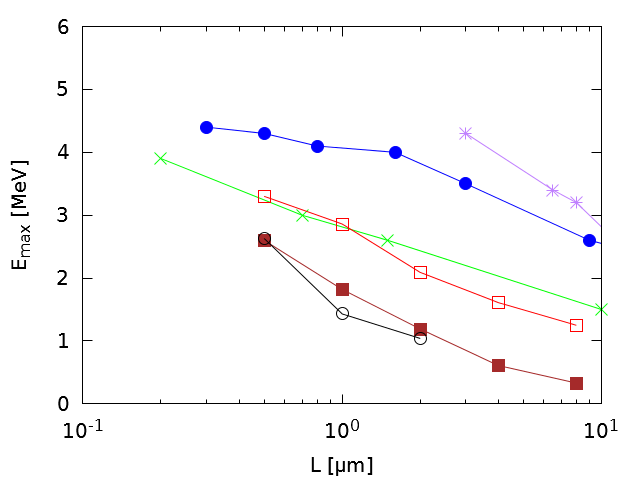}
\caption{Plot of $\Emax$ versus $L$ in logarithmic scale 
from various experiments with a laser pulse having $a_0\sim 3$
and a metal target: Ceccotti experiment (45$^\circ$ incidence angle) from ref. \cite{Ceccotti} (blue circles),
Neely experiment (30$^\circ$) from ref. \cite{Neely} (green crosses), Flacco experiment (45$^\circ$) from ref. \cite{Flacco} (purple stars). These data are compared with the results of
our 2D PIC simulation at zero degree incidence (filled red squares), 2D at 30$^\circ$ incidence (empty red squares) and 3D PIC simulation at zero degree incidence (empty black circles).}
\label{fig:7} 
\end{figure}

The comparison with the experimental results is a challenging task: in figure \ref{fig:7}, we
show the results of some experiments whose laser pulse has the same P-polarization,
with a duration and intensity very close to the ones considered here,
and whose target has the same structure, namely a metal foil plus contaminants.
The cut-off energy increases as the target thickness is reduced, until the effect of
finite contrast prevails inverting the trend. The results of various experiments
differ by more than a factor two, but the decreasing trend is similar and the same behaviour can be
seen in the 2D and 3D PIC simulations.
%
%

\begin{table}[ht]
\centering 
\begin{tabular}{c c c c c c }
\hline \hline 
$L$ & $\Emax(ct=50)$ & $E_\infty^{(2D)}$ & $ct^{*(2D)}$ & $\sigma_E$ & $\sigma_{ct^*} $ \\ [0.5ex] 
\hline 
 0.5 & 2.64 & 2.62 & 17.5 & 0.05 & 0.03 \\
 1 & 1.82 & 1.82 & 18.0 & 0.02 & 0.15 \\
 2 & 1.19 & 1.19 & 18.4 & 0.02 & 0.2 \\
 4 & 0.58 & 0.61 & 19.9 & 0.02 & 0.5 \\
 8 & 0.25 & 0.33 & 23.3 & 0.02 & 0.9 \\
 \\ [1ex] 
\hline 
\end{tabular}
\caption{Fitting parameters for 2D simulations with zero incidence angle and
 target thicknesses $0.5\leq L\leq 8 \mu$m.
 The chosen interval for fitting is $ct_1=20 \mu$m and $ct_2=80 \mu$m }
\label{tab:1}
\end{table}

\begin{table}[ht]
\centering 
\begin{tabular}{c c c c c c }
\hline \hline 
$\alpha$ & $\Emax(ct=50)$ & $E_\infty^{(2D)}$ & $ct^{*(2D)}$ & $\sigma_E$ & $\sigma_{ct^*} $ \\ [0.5ex] 
\hline 
 5 & 1.28 & 1.40 & 19.9 & 0.01 & 0.1 \\
 10 & 1.47 & 1.62 & 20.1 & 0.01 & 0.1 \\
 15 & 1.59 & 1.82 & 20.7 & 0.0215 & 0.15 \\
 \\ [1ex] 
\hline 
\end{tabular}
\caption{Fitting parameters for 2D simulations for three different incidence angles $\alpha=5^{\circ},10^{\circ},15^{\circ}$
 and target thickness $L=2 \mu$m.
 The chosen interval for fitting is $ct_1=20 \mu$m and $ct_2=80 \mu$m and the fitting errors are quoted. }
\label{tab:2}
\end{table}

\
\begin{table}[ht]
\centering 
\begin{tabular}{c c c c c c }
\hline \hline 
$ L $ & $\Emax(ct=50)$ & $E_\infty^{(3D)}$ & $ct^{*(3D)}$ & $\sigma_E$ & $\sigma_{ct^*} $ \\ [0.5ex] 
\hline 
 0.5 & 1.25 & 2.63 & 15.3 & 0.01 & 0.2 \\
 1 & 0.56 & 1.43 & 18.9 & 0.02 & 0.1 \\
 2 & 0.44 & 1.04 & 17.3 & 0.01 & 0.1 \\
 \\ [1ex] 
\hline 
\end{tabular}
\caption{Fitting parameters for 3D simulations for zero incidence angle and three different target
 thicknesses $L=0.5,1,2 \mu$m.
 The chosen interval for fitting is $ct_1=20 \mu$m and $ct_2=60 \mu$m and the fitting errors are quoted. }
\label{tab:3}
\end{table}

\section{Conclusions}
The asymptotic value of the cut-off energy of protons, which is what is measured in 
experiments, is difficult to extract from PIC simulations.
Indeed, the 2D results do not exhibit a saturation, whereas the 3D results show that
a saturation might be reached, despite at a large time ($ct > 200 \mu$m), which is 
computationally too expensive to be reached. We propose here a simple recipe based on the Schreiber et al. \cite{Schreiber}
model, which assumes that the acceleration of protons present in the contaminants is due to
the positive surface charge created on the rear target, thanks to the escape of the electrons.
In the 3D version, the charged spot is circular with a radius $R$ comparable with laser
waist. The rise in time of the cut-off energy can be analytically computed.
We have formulated an analogous 2D model where the charge is on an infinite strip of height $2R$
and we obtain a simple asymptotic expression for the rise in time of the cut-off energy,
which does not saturate but exhibits a logarithmic growth, just as in 1D models
of the vacuum expansion of plasma. The analytical results suggest two phenomenological laws,
which depend on the asymptotic energy $E_\infty$ and the time $t^*$ at which the acceleration
begins. 
The fits to the 2D and 3D results coming from PIC simulations are quite good and the statistical uncertainties $\sigma_{E_\infty}/E_\infty$
and $\sigma_{ct^*}/ct^*$ are quite small (a few percent). The extrapolated values $E_\infty^{(2D)}$ and
$E_\infty^{(3D)}$, computed for different target thickness, are comparable and moreover they
can be obtained from the results with $ct\le 50\sim 60 \mu$m, which is reachable also in 3D numerical simulations.
The fitting appears to be satisfactory also for small incidence angles, even though the model was developed for
normal incidence. 

To conclude, we believe that, for the targets that we analysed, in which the protons are only on the
thin layer above the bulk, the proposed phenomenological model is adequate to avoid the
arbitrariness in the choice of the time at which the asymptotic cut-off energy is chosen usually in numerical simulations. 
In addition, the parametric explorations, which can be carried out only in 2D, may have
a quantitative value, with an adequate extrapolation, rather than being of purely qualitative nature.
The results we have presented refer to a specific intensity and a range of target thicknesses
chosen in order to fulfil the applicability conditions of the model.
%
%
%
%
%
%
%
%

\section*{Acknowledgements} The work has been done within the L3IA INFN Collaboration, which the authors would like to thank all.

\section*{References}

\newpage
\section{Appendix}
Let's consider a target which is infinitely extended along the plane $xy$ and delimited by the planes $z=-L$ and $z=0$. 
We can consider a circular radius $r_L$ which we assume to be the spot of the laser 
pulse propagating along $z$. The electrons are heated and diffused by the laser itself. Supposing 
that they diverge with angle $\theta$, the electrons will leave the plane $z=0$ from a disc of radius 
\[ R= r_L + L \tan \theta \]
We assume that the target is a metallic foil and that the protons are in the contaminants deposited on the plane $z=0$.
The electrons are heated, diffuse and cross the $z=0$ boundary leaving the target and inducing on it a positive
charge density $\sigma(t)$, that we suppose varies slowly with $t$. If $Qe$ is the total number of positive charge
on the surface, the density is
\begin{equation}
\sigma= { Q e\over \pi \,R^2}
\label{eq:3d_density}
\end{equation}
This is the geometry for the 3D case, that we shall treat analytically

We consider another geometry in which the electrons on the plane $z=L$ leave the rectangle $|x|\leq R$, $|y|\leq L$ of area $4LR$ .
In this case the density is given by 
\begin{equation}
\sigma= { Q e\over 4 RL}
\label{eq:2d_density}
\end{equation}
and we may assume that the laser spot on $z=0$ is $|x|\leq R$ and $|y|\leq L$.
The intensity defined as the power per unit surface is assumed to be the same for
both geometries.

\subsection{The 3D case: charge density on a disk}
Using cylindrical coordinates and computing the potential corresponding to the surface density \ref{eq:3d_density}
\[
\begin{split}
V(z) &= 2\pi\sigma \,\int_0^R \,r dr {1\over \sqrt{r^2+z^2}}= \pi \sigma \int_0^R \,dr^2 {1\over \sqrt{ z^2+r^2}}= \\
&= 2\pi \sigma [ \sqrt{z^2+R^2} -z]
\end{split}
\]
Introducing the dimensionless variable $\zeta= z/R$ we have
\[ V(\zeta)= 2\pi\,R \sigma[ \sqrt{1+\zeta^2} -\zeta ]\]
Since $V(0)= 2\pi\,R\,\sigma$ we redefine the potential by subtracting it.
\begin{equation}
\hat{V}(\zeta)= V(\zeta) - V(0) = 2\pi\,R \sigma[ \sqrt{1+\zeta^2} -\zeta -1]
\label{eq:potential}
\end{equation}
The potential energy is given by $eV(\zeta)$. We notice that we have 
\[
\left\{
  \begin{array}{ll}
    \displaystyle \hat{V}(z) \simeq -2\pi \sigma \,z              &\mathrm{for}\; z\to 0 \\
    \displaystyle \hat{V}(z) \simeq {eQ\over z} - {2Qe \over R}   &\mathrm{for}\; z\to \infty
  \end{array}
\right.
\]

Letting $v=\dot z$ and assuming $v(0)=0$, namely that the protons are initially at
rest on the surface $z=0$, we can apply the energy conservation
\[ m{v^2\over 2}+eV(\zeta)\equiv E+eV(\zeta)=0 \]
Calling $v_\infty$ the speed reached at infinite distance
\[ E_\infty= m{v_\infty ^2\over 2} = -eV(\infty)= {2Qe^2\over R} = 2\pi \,e \, R\, \sigma \]
we can define
\[ -eV(\zeta)= 2\pi\,e\, R\sigma \,s(\zeta) = m{v_\infty ^2\over 2} s(\zeta) \]
where from equation \ref{eq:potential}
\[ s(\zeta)= 1+\zeta-\sqrt{1+\zeta^2} \]
As a consequence we have 
\begin{equation}
E= E_\infty s(\zeta) \qquad \qquad v=v_\infty \sqrt{s(\zeta)}
\end{equation}
We introduce the new variables
\[ X=\sqrt{s} \qquad \qquad \tau=t{ v_\infty \over R} \]
Then we have
\begin{equation} \label{eq:5}
{d\zeta\over d\tau}= {v\over v_\infty}= \sqrt{s(\zeta)}
\end{equation}
We might solve this equation with initial condition $\zeta(0)=0$. We rather solve the equation for $X$
\begin{equation} \label{eq:6}
{dX\over d\tau}= {dX\over ds}\,\,{ds\over d\zeta}\,\,{d\zeta\over d\tau}={1\over 2}\,{ds\over d\zeta}
\end{equation}
Let us notice that
\[ {dX\over d\tau } = {1\over 2}\parton{1-{\zeta \over \sqrt{1+\zeta^2}}} = {1\over 2}\parton{ 1 +{\zeta\over 1-s}}^{-1} \]
inverting $s=s(\zeta)$ we have $\zeta= (2s-s^2)/(2(1-s))$ and finally replacing in the r.h.s. of the last equation we obtain 
\[ {dX\over d\tau } = \parton{ 1 +{1\over (1-s)^2}}^{-1} = \parton{ 1 +{1 \over (1-X^2)^2 }}^{-1} \]
The results is obtained with integration by parts
\[
\begin{split}
\tau &= X + \int_0^ X {du\over (1 - u^2)^2} = \left. X - {1\over 2}{d\over d\alpha} \,\int_0^X \,{1\over \alpha^2-u^2} \right |_{\alpha=1} = \\
&= X + {1\over 2}\,\, {X\over 1-X^2} + {1\over 4}\, \log{1+X\over 1-X}
\end{split}
\]
Asymptotically, for $\tau\to \infty$, we have $X\to 1$
\[ \tau \sim {1\over 4(1-X) } \qquad \qquad X\simeq 1-{1\over 4\tau} \]
The energy asymptotic behaviour is given by $E/E_\infty= s=X^2$ and consequently for $t\to \infty$
\[ E\simeq E_\infty\,\parton{ 1-{1\over 4\tau}}^2 \]
\subsection{The 2D case: charge on slab}
We consider the slab $|x|\leq R$ and $|y|\leq L$ on the rear surface $z=0$ where the density is given by \ref{eq:2d_density}. The potential is given by
\begin{equation} \label{eq:2d_potential}
\begin{split}
V(z) &= \sigma \int_{-R}^R \,dx \int_{-L}^L\,{dy \over \sqrt{x^2+y^2+z^2}} = \\
&= 4 \sigma \int_0^R \,dx \int_0^{L/\sqrt{x^2+z^2}}\,{du\over \sqrt{1+u^2}} = \\
&= 4\sigma \, \int_0^R \,dx\,\mathrm{arsinh}\parton{L\over \sqrt{x^2+z^2}}
\end{split}
\end{equation}
Since $4\sigma= eQ/(LR)$ we first consider the limit $L\to 0$ which corresponds to the density $\sigma(z)= {eQ/(2R)}\delta(y)$ and the result is
\[ V(z)= {eQ\over R}\,\, \int_0^R \,dx\,\parton{1\over \sqrt{x^2+z^2}} = {eQ\over R}\,\mathrm{arsinh}{1\over \zeta}\qquad \qquad \zeta={z\over R} \]
Recalling that $\mathrm{arsinh}(u)= \log(u+\sqrt{1+u^2})$ we see that $V(\zeta)\sim \log(2/\zeta)$ for $\zeta \to 0$ whereas it vanishes as $1/\zeta$ for $\zeta \to \infty$. As a consequence we cannot have $V$ vanishing at $\zeta=0$ with a subtraction. Indeed if we compute $V(0)$ we see that it diverges as $\log (1/L)$ for $L\to 0$ (see eq. \ref{eq:diverging_potential_as_log}).
We wish to define a potential which vanishes at $z=0$ as a consequence in the definition we have to subtract $V(0)$. This can be done for any finite value of $L$ and also for $L\to \infty$. In order to compute $V(0)$ for a given non vanishing $L$ we set $\xi=x/L$ and integrating by parts we obtain
\begin{equation} \label{eq:10}
\begin{split}
V(0) &= {eQ\over R}\,\int_0^{R/L} \,d\xi \,\mathrm{arsinh}{1\over \xi} \\
&= {eQ\over R} \parqua{ \left. \xi\mathrm{arsinh}{1\over \xi}\right |_0^{R/L} +\int_0^{R/L} {d\xi\over \sqrt{1+\xi^2}} } = \\
&= {eQ\over R} \parqua{ {R\over L} \mathrm{arsinh}{L\over R}+ \mathrm{arsinh}{R\over L} }
\end{split}
\end{equation}
We see that $V(0)$ is finite for any $L>0$, that it diverges as $ \log(1/L)$ for $L\to 0$ and that it vanishes for $L\to \infty$.
We redefine the potential as 
\[
\begin{split}
\hat V(z) &= V(z)-V(0) = \\
&={eQ\over RL}\, \, \int_0^R {\,dx\, \parqua{ \mathrm{arsinh}\parton{L\over \sqrt{x^2+z^2}} -\mathrm{arsinh}{L\over x} }}
\end{split}
\]
Let us consider the asymptotic behaviour of $V(z)$ for $z\to \infty$ for $L$ having any fixed finite value. To this end, we recall that when $ u=L/\sqrt{x^2+z^2} \to 0$
we can approximate $\mathrm{arsinh}$ with its Taylor expansion $\mathrm{arsinh}(u)= u-u^3/6+O(u^5)$ retaining only the first term we have
\[ V(z)= {eQ\over R}\int_0^{R/z}\,{du\over \sqrt{1+u^2}} = {eQ\over R}\,\mathrm{arsinh}{R\over z} \simeq {eQ\over z} \]
We consider now the limit $L\to \infty$. In this limit it is evident that $V(0)=0$. Moreover, starting from equation \ref{eq:2d_potential} and computing the electric field, we have
\[
\begin{split}
\Ecal_z &=-\derp{V}{z}= 4\sigma \int_0^R {dx\,{1\over \sqrt{1+{L^2\over x^2+z^2}} }\,{L\,z\over (x^2+z^2)^{3/2}} }= \\
&=4\sigma \int_0^R {{dx\over z}\,{ 1\over 1+{ \displaystyle x^2\over \displaystyle z^2}} \,{1\over \parton{1+{ \displaystyle x^2+z^2\over \displaystyle L^2}}^{1/2}}}
\end{split}
\]
If we take the limit for $L\to \infty$ we recover the following result
\begin{equation}
\Ecal_z = 4\sigma \arctan{R\over z} \qquad \qquad \Ecal_z \sim { 4\sigma R \over z} \qquad \mathrm{for}\; z\to \infty
\label{eq:2d_electric_field}
\end{equation}
As a consequence the potential behaves as $V(z) \simeq - 4\sigma R \log(R/z)$ for $z\to \infty$.
We compute exactly the potential corresponding to \ref{eq:2d_electric_field} introducing again the dimensionless variable $\zeta=z/R$
\begin{equation} \label{eq:diverging_potential_as_log}
\begin{split}
V(z) &=- 4\sigma \int_0^z {\arctan{R\over z'}\,dz'} = \\
&= 4R\sigma \parton{ -\zeta\arctan {1\over \zeta} +\log{1\over \sqrt{\left(1+\zeta^2\right)}} }
\end{split}
\end{equation}
where manifestly $V(0)=0$.

The potential now diverges for $z\to \infty$ but we still use the energy conservation 
\[ E+eV=0 \qquad E= -eV= E_\infty s(\zeta) \]
where we put, in analogy with the 3D, 
\[
\begin{split}
E_\infty &\equiv m{v^2_\infty\over 2}= 4\,e\, R\sigma \\
s(\zeta) &=\zeta\arctan {1\over \zeta} - \log{1\over \sqrt{1+\zeta^2}}
\end{split}
\]
and the equation \ref{eq:5} holds for the coordinate $\zeta$. As in the 3D case we introduce the coordinate $X=\sqrt{s}$ and equation \ref{eq:6} holds. In order to simplify the analysis we replace $s(\zeta)$ defined by \ref{eq:10} with
$s(\zeta)= \log(1+\zeta)$ which has the same asymptotic behaviour at $\zeta=0$ and $\zeta\to \infty$. Finally we have 
\[ {dX\over d\tau} ={1\over 2}{1\over 1+\zeta}= {e^{-s}\over 2} = {1\over 2}\,e^{-X^2} \]

The solution reads
\[\tau= 2\int_0^X e^{u^2}\,du= e^{x^2}\parqua{ {1\over x} + {1\over 2x^3} +O\parton{1\over x^5}} \]
retaining only the first term we invert the equation
\[ x^2=\log \tau +\log x \qquad \qquad x^2= \log\tau + {1\over 2} \log \log \tau +\ldots \]
The results is given by 
\[ E= E_\infty \parqua {\log \tau + {1\over 2} \log \log \tau } \qquad \qquad \tau=t {v_\infty\over R} \]

\end{document}